\documentclass[reprint, superscriptaddress, amsmath,amssymb, aps,pra,]{revtex4-1}

\def\PRA{{Phys.~Rev.~A} }
\def\Adv{{Adv.~At.~Mol.~Opt.~Phys.} }
\def\PRB{{Phys.~Rev.~B} }

\def\Natc{{Nat. Commun.} }
\def\Sci{{Science}}
\def\Nat{{Nature} }
\def\Natp{{Nat. Phys.} }

\def\JPB{{J.~Phys.~B} }
\def\PRL{{Phys.~Rev.~Lett.} }

\def\PCCP{{Phys. Chem. Chem. Phys.}}
\def\CPc{{Comput. Phys. Commun.} }
\def\OPTC{{Opt. Commun.}}

\usepackage{graphicx}
\usepackage{dcolumn}
\usepackage{bm}
\usepackage{siunitx}
\DeclareSIUnit\atomicunit{a.u.}

\usepackage{braket}
\usepackage{color}

\begin{document}
\title{Exploring Valence Electron Dynamics of Xenon through Laser-Induced Electron Diffraction}

\author{Fang Liu}
\email{fang.liu@uni-jena.de}
\affiliation{Helmholtz-Institut Jena, 07743 Jena, Germany}

\affiliation{Theoretisch-Physikalisches Institut, Friedrich-Schiller-Universit\"at Jena, 07743 Jena, Germany}

\affiliation{GSI Helmholtzzentrum f\"ur Schwerionenforschung GmbH, 64291 Darmstadt, Germany}

\author{Slawomir Skruszewicz}
\affiliation{University of Hamburg, Germany}

\author{Julian Späthe}
\affiliation{Institute of Optics and Quantum Electronics, 07743 Jena, Germany}

\author{Yinyu Zhang}
\affiliation{Institute of Optics and Quantum Electronics, 07743 Jena, Germany}

\author{Sebastian Hell}
\affiliation{Institute of Optics and Quantum Electronics, 07743 Jena, Germany}

\author{Bo Ying}
\affiliation{Institute of Optics and Quantum Electronics, 07743 Jena, Germany}
\affiliation{Helmholtz-Institut Jena, 07743 Jena, Germany}

\author{Gerhard G. Paulus}
\affiliation{Institute of Optics and Quantum Electronics, 07743 Jena, Germany}
\affiliation{Helmholtz-Institut Jena, 07743 Jena, Germany}

\author{Bálint Kiss}
\affiliation{ELI-ALPS, ELI-HU Non-Profit Ltd., Wolfgang Sandner utca 3., Szeged, H-6728, Hungary}

\author{Krishna Murari}
\affiliation{ELI-ALPS, ELI-HU Non-Profit Ltd., Wolfgang Sandner utca 3., Szeged, H-6728, Hungary}

\author{Malin Khalil}
\affiliation{ELI-ALPS, ELI-HU Non-Profit Ltd., Wolfgang Sandner utca 3., Szeged, H-6728, Hungary}

\author{Eric Cormier}
\affiliation{ELI-ALPS, ELI-HU Non-Profit Ltd., Wolfgang Sandner utca 3., Szeged, H-6728, Hungary}

\author{Li Guang Jiao}
\affiliation{Theoretisch-Physikalisches Institut, Friedrich-Schiller-Universit\"at Jena, 07743 Jena, Germany}

\affiliation{College of Physics, Jilin University, Changchun 130012, People's Republic of China}

\author{Stephan Fritzsche}
\affiliation{Helmholtz-Institut Jena, 07743 Jena, Germany}

\affiliation{Theoretisch-Physikalisches Institut, Friedrich-Schiller-Universit\"at Jena, 07743 Jena, Germany}

\affiliation{GSI Helmholtzzentrum f\"ur Schwerionenforschung GmbH, 64291 Darmstadt, Germany}

\author{Matthias K\"{u}bel}
\email{matthias.kuebel@uni-jena.de}

\affiliation{Helmholtz-Institut Jena, 07743 Jena, Germany}
\affiliation{Institute of Optics and Quantum Electronics, 07743 Jena, Germany}

\date{\today}

\begin{abstract}


Strong-field ionization can induce electron motion in both the continuum and the valence shell of the parent ion. Here, we explore their interplay by studying laser-induced electron diffraction (LIED) patterns arising from interaction with the potentials of two-hole states of the xenon cation.  The quantitative rescattering theory is used to calculate the corresponding photoelectron momentum distributions, providing evidence that the spin-orbit dynamics could be detected by LIED. We identify the contribution of these time-evolving hole states to the angular distribution of the rescattered electrons, particularly noting a distinct change along the backward scattering angles. We benchmark numerical results with experiments using ultrabroad and femtosecond laser pulses centered at \SI{3100}{nm}.


\end{abstract}


\maketitle

\section{Introduction}

Laser-induced electron diffraction (LIED) has been established as a powerful alternative to conventional electron diffraction, see Refs.~\cite{Giovanni2023,Amini2020} for recent reviews. The technique relies on the laser-driven elastic rescattering \cite{corkum} of a photoelectron emitted by strong-field ionization, which gives rise to high-order above threshold ionization (HATI), the basic strong-field phenomenon underlying LIED. The rescattering process provides LIED with two interesting properties: first, an ultrahigh current density, allowing imaging on the single molecule level \cite{Meckel2008}; second, perfect synchronization between ionization and scattering events, allowing (attosecond) time-resolved experiments \cite{Niikura2002,Blaga2012,Wolter2016}. 

The LIED signal can be described by the atomic scattering cross sections and a molecular interference term \cite{BWalker1996, Lein2002, Meckel2008}. Numerous LIED experiments have focused on the measurement of the molecular interference term and the accurate, and time-dependent, retrieval of molecular bond lengths has been demonstrated \cite{Blaga2012, Wolter2016, Sanchez2021}. Notably, these measurements require the recolliding electron to possess a sufficiently short de Broglie wavelength and correspondingly large momentum. For this reason, long driving wavelengths $(\lambda \gtrsim \SI{2}{\micro m})$ are favorable to drive LIED experiments for molecular structure retrieval. In the case of atoms or lower recollision momentum, the scattering signal is governed by the differential elastic electron scattering cross-sections (DCS), which depend on the valence electron distribution \cite{QRS_HATI2}. Good agreement between measured LIED patterns and DCS known from conventional electron diffraction experiments has been obtained \cite{Sultana1987, Ireneusz2006}. 

Specifically, strong-field driven tunnel ionization is not only the primary step in HATI and LIED but can also initiate electronic and nuclear dynamics inside the parent ion, thus acting as a ``pump''. This has enabled ultra-stable pump-probe type experiments: attosecond time resolution is obtained by exploiting the perfect synchronization between the laser field and the recolliding electron \cite{corkum}. As the returning electron wave packet is chirped, an energy-resolved measurement of the returning electron, acting as a probe, provides access to different pump-probe delays. This principle has allowed for time-resolved measurements of nuclear dynamics in high-harmonic spectroscopy \cite{Baker2006, Shafir2012} and LIED experiments \cite{Blaga2012, Wolter2016}. While the prospect of employing high-harmonic generation (HHG) for probing \cite{Baker2006, Haessler2010} is inherently appealing, a significant challenge arises from the strong impact of phase matching on HHG \cite{Nalda2004, Lappas2000, Shiner2011}. In contrast, LIED is insensitive to phase matching, representing a promising alternative. However, to the best of our knowledge, LIED experiments revealing electron-hole dynamics have not yet been reported. 

Here, we study the interplay of continuum and bound electron dynamics in the HATI process. Our approach is best illustrated by viewing laser-induced recollision as a pump-probe experiment \cite{Ichibayashi2009, Ichibayashi2011}:
tunnel ionization takes the role of a pump pulse which essentially starts two clocks: a laser-dependent one that corresponds to the field-driven motion of free electrons undergoing elastic re-scattering, and a target-dependent one that relates to the bound electron dynamics. Both clocks are read at the time of recollision, when the recolliding electron probes the hole density of the ion by elastic scattering. 

In this work, we consider the Xe atom. The ground state of the Xe cation has two fine-structure components: the $^2P_{3/2}$ and $^2P_{1/2}$ that are separated by $\Delta{E_{\mathrm{SO}}}=1.3$ eV due to the spin-orbit interaction. Both ion states are coherently populated by tunnel ionization, thus creating a wave packet. As the spin-orbit wave packet evolves, the $5p^5$ electron-hole (vacancy) in the valence shell oscillates between the $m = 0$ state and the $|m| = 1$ states of the valence shell of the Xe$^+$ ion with the period $T_{\mathrm{SO}}$ = $h/ \Delta{E_{\mathrm{SO}}}$ ($3.2$ fs), where $m$ is the magnetic quantum number~\cite{JPB}. The spatial hole density in the valence shell is described by the orbitals for $m = 0$ (``peanut shape") and $|m| = 1$ (``donut shape"). At integer $n$ and half-integer $\left(n+\frac{1}{2}\right)$ multiples of the spin-orbit period(where $n=0,1,2,3,\ldots$), the hole alternately populates the $m=0$ and $|m|=1$ orbitals, respectively. The oscillating hole density has been tracked in Kr using attosecond transient spectroscopy \cite{M15}. For Ne and Ar ion momentum spectroscopy \cite{Fleischer2011} or momentum imaging of direct electrons \cite{Fechner2014, M} has been applied. Recently, the spin-orbit wave packet in Xe has been probed using sequential double ionization in an elliptically polarized near-infrared laser field \cite{Steward2023}. Here, we employ elastic rescattering in a mid-infrared field ($\lambda = \SI{3100}{nm}$) with an optical period of $T = \SI{10.5}{fs}$. Owing to the relatively long optical period, the returning electron wave packet spans several femtoseconds, allowing us, in principle, to probe the evolution of the spin-orbit wave packet in xenon. 

The article is structured as follows. In Sec.~\ref{Sec_TheoreticalFramework}, with the reconstructed electron-hole potential, we introduce the quantitative rescattering theory (QRS) model used to calculate the photoelectron momentum distributions (PMDs) for HATI. Based on QRS model, the simulated results are shown and
discussed in Sec.~\ref{Sec_Results}. Finally, Sec.~\ref{Sec_Conclusions} contains conclusions and outlook.

Unless indicated otherwise, atomic units ($m_e = e = \hbar = 4\pi\varepsilon_0 = 1$) are used throughout the paper.

\section{Theoretical model} \label{Sec_TheoreticalFramework}

\subsection{The strong-field approximation}

In the strong-field approximation (SFA), the first two terms of the perturbation series, called direct (SFA1) and rescattering (SFA2) amplitudes, respectively, express the momentum-dependent ionization amplitude as: 
\begin{eqnarray} \label{SFA}
    f_\mathrm{\,SFA} (\bm{p}) = f_\mathrm{\,SFA1} (\bm{p}) + f_\mathrm{\,SFA2} (\bm{p}),
\end{eqnarray}
where $\bm{p}$ is the momentum of the detected photoelectron. The direct ionization amplitude in Eq.~(\ref{SFA}) is given by~\cite{RWPR},
\begin{eqnarray}\label{SFA1}
    f_\mathrm{\,SFA1} (\bm{p}) = -i \int _{-\infty}^{\infty} dt \bra{\chi_{\bm{p}}(t)} \bm{r} \cdot \bm{F}(t) \ket{\Psi_i(t)},
\end{eqnarray}
where $\bm{F}(t)=- \partial \bm{A} (t)/\partial t$ is the laser electric field, and $\Psi_i(t)$ is the initial ground state wave function. The Volkov state $\chi_{\bm{p}}(t)$ in Eq.~(\ref{SFA1}) is given by
\begin{eqnarray} \label{Volkov}
    \bra{\bm{r}}\chi_{\bm{p}}(t)\rangle= \frac{1} {(2 \pi)^{3/2}} e^{i[\bm{p}+\bm{A}(t)] \cdot \bm{r}} e^{-iS(\bm{p},t)},
\end{eqnarray}
where the action $S$ is given by
\begin{eqnarray} \label{Sp}
    S(\bm{p},t) = \frac{1} {2} \int _{-\infty}^{t}  dt^{\prime} [\bm{p}+\bm{A}(t^{\prime})]^{2}.
\end{eqnarray}

The second term in Eq.~(\ref{SFA}), the so-called rescattering amplitude, accounts for laser-induced elastic scattering of the returning
electron from the parent ion. This rescattering amplitude can be expressed as:
\begin{eqnarray}
\label{SFA2}
   f_\mathrm{\,SFA2}(\bm{p})=&-\int_{-\infty}^{\infty}dt\int_{t}^{\infty}dt^{\prime}\int
   d\bm{k} \bra{\chi_{\bm{p}}(t^{\prime})}V \ket{\chi_{\bm{k}}(t^{\prime})} \nonumber \\
   & \times \bra{\chi_{\bm{k}}(t)}\bm{r} \cdot \bm{F}(t)\ket{\Psi_i(t)},
\end{eqnarray}
where $V$ is the scattering potential. It takes the form
\begin{eqnarray} \label{V}
    V(r) = \widetilde{V}(r) e^{- \alpha r},
\end{eqnarray}
where $\alpha$ is a screening factor introduced to avoid the singularity in the integrand in Eq.~(\ref{SFA2}) and $\widetilde{V}(r)$ is the atomic model potential that can be written in the form
\begin{eqnarray} \label{V_pot}
    \widetilde{V}(r)=-\frac{1+a_1e^{-a_2r}+a_3r e^{-a_4r}+a_5e^{-a_6r}}{r}.
\end{eqnarray}
The parameters $a_i (i=1,3,5)$ can be found in Ref.~\cite{XMTong}. As can be seen from Eq.~(\ref{SFA2}), the rescattering amplitude consists of three time-ordered steps
by the electron: the initial tunnel ionization, propagation in the laser field as well as elastic scattering with the parent ion.

\subsection{Elastic differential cross sections}

In this section, we briefly summarize the standard potential scattering theory which has been well documented in the textbook Ref.~\cite{book}. The scattered wavefunction of an electron by a spherical potential $V(r)$ satisfies the time-independent Schr\"{o}dinger equation
\begin{eqnarray} \label{TDSE}
[\nabla^2 +k^2 - U(r)]\mathbf{\psi}(\bm{r}) =0,
\end{eqnarray}
where $U(r) = 2V(r)$ is the reduced potential and $k$ is the electron momentum, related to the electron energy by $k = \sqrt{2E}$. For a short-range potential which falls faster than $r^{-2}$ as $r \rightarrow \infty$, the wavefunction of the scattered electron in the asymptotic region is given by
\begin{eqnarray}\label{asymptoticwave}
    \mathbb{\psi}^+(\bm{r})_{r\rightarrow \infty} = \frac{1}{(2\pi)^{3/2}}[e^{i\bm{k} \cdot \bm{r}}+f(\theta)\frac{e^{ikr}}{r}],
\end{eqnarray}
where $f(\theta)$ is the scattering amplitude and $\theta$ is the polar angle measured from the incident direction.

To obtain the scattering amplitude, we solve Eq.~(\ref{TDSE}) by expanding the scattered wavefunction in partial waves,
\begin{eqnarray}\label{Eq_expandingwave}
\mathbb{\psi}^+(\bm{r}) = \sqrt{\frac{2}{\pi}} \frac{1}{kr}\sum_{lm} i^{l} u_l(k,r)Y_{l m}(\hat{\bm{r}})Y^*_{l m}(\hat{\bm{k}}), 
\end{eqnarray}
where $Y_{lm}$ is a spherical harmonic. The continuum waves are
normalized to $\delta(\bm{k}-\bm{k^{\prime}})$. The radial equation $u_l(k, r)$ satisfies
\begin{eqnarray}\label{RW}
[\frac{d^2}{dr^2} + k^2 - \frac{l(l+1)}{r^2} - U(r)]u_l(k,r) =0.
\end{eqnarray}

For a plane wave when $V(r) = 0$, the radial component $u_l(k, r)/kr$ in Eq.~(\ref{Eq_expandingwave}) is a standard spherical Bessel function $j_l(kr)$.

When $r\rightarrow \infty$, the boundary condition satisfied by $u_l(k, r)$
for $V(r) = 0$ is
\begin{eqnarray}\label{Eq_Radialwave1}
u_l(k, r) = \sin(kr-\frac{1}{2}l\pi),
\end{eqnarray}
while for a short-range potential $V(r)$,
\begin{eqnarray}\label{Eq_Radialwave2}
u_l(k, r) = e^{({i\delta_l})}\sin(kr-\frac{1}{2}l\pi+\delta_l), 
\end{eqnarray}
 where $\delta_l$ is the phase shift, that displays the influence of the interaction.

By matching the coefficients of the outgoing spherical waves in Eqs. (\ref{asymptoticwave}) and (\ref{Eq_expandingwave}), and using Eqs. (\ref{Eq_Radialwave1}) and (\ref{Eq_Radialwave2}), the scattering amplitude is given by
\begin{eqnarray}\label{hamiltonian}
    f(\theta) =\sum_{l=0}^{\infty}  \frac{2l+1}{k} e^{\mathrm{i}{\delta_l}}\sin(\delta_l)P_l(\cos\theta),
\end{eqnarray}
where $P_l(\cos\theta)$ are the Legendre polynomials.

For the scattering by a Coulomb potential is given by
\begin{eqnarray}\label{pureCoulomb}
    V_c(r) = \frac{Z_1Z_2}{r},
\end{eqnarray}
where $Z_1$ and $Z_2$ are the charges of the projectile and the target, respectively. Since the Coulomb interaction drops off so slowly for large $r$, it can be treated in parabolic coordinates and the scattering amplitude can be obtained analytically
\begin{eqnarray}\label{AmppureCoulomb}
    f_c(\theta) =- \eta e^{2\mathrm{i}{\sigma_0}}  \frac{e^{-\mathrm{i \eta} \mathrm{ln}[\sin^2(\theta/2)]}}{2k\sin^2(\theta/2)},
\end{eqnarray}
where 
\begin{eqnarray}\label{hamiltonian}
    \sigma_0 =- \mathrm{arg}[\Gamma(1+\mathrm{i \eta})] \, , \,
    \eta = \frac{Z_1Z_2}{k}.
\end{eqnarray}

In order to mimic the partial screening of the nuclear charge by the electrons, a short-range potential $V(r)$ is added to a Coulomb potential $V_c(r)$, using partial-wave expansion, the scattering amplitude can be expressed by
\begin{eqnarray}\label{AmpCoulomb}
    \hat{f}(\theta) =\sum_{l=0}^{\infty}  \frac{2l+1}{k} e^{2\mathrm{i}{\sigma_l}} e^{\mathrm{i}{\delta_l}}\sin(\delta_l)P_l(\cos\theta) ,
\end{eqnarray}

Thus, the scattering amplitude for the general case is given by
\begin{eqnarray}\label{FullAmp}
    f(\theta) =f_c(\theta)+\hat{f}(\theta) ,
\end{eqnarray}
and the elastic scattering DCS for a given energy
reads
\begin{eqnarray}\label{DCS}
    \frac{d\sigma_{el}(k,\theta)}{d \Omega_r} =  |f(\theta)|^2.
\end{eqnarray}

\subsection{The electron-hole potential}

Here the scattering
potentials used in the numerical calculations are given. We consider the elastic scattering of electrons with the Xe$^+$ ion. The DCSs for the $m = 0$ and  $|m| = 1$ vacancy states are calculated using standard potential scattering theory. 

The static potential $V(r)$ of the Xe$^+$ ion is structured as, 
\begin{eqnarray}\label{atomicpotential}
     V(\bm{r})  =  -\frac{Z}{r} + V^{\mathrm{DFS}}(r) -V_{1 m_{0,1}}(\bm{r}),
\end{eqnarray}
where $Z$ is the nuclear charge of the target, $V^{\mathrm{DFS}}(r)$ is the Dirac-Fock-Slater potential where the summation runs over all orbitals (electrons). The term $V_{1 m_{0,1}}(\bm{r})$ is a hole potential that describes the Coulomb interaction between projectile electron and the orbital $1 m_{0,1}$ in the ion. It is given by:
\begin{eqnarray}\label{holepotential}
     V_{1 m_{0,1}}(\bm{r})= \int |\psi_{1 m_{0,1}}(\bm{r}^\prime)|^2 \frac{1}{|\bm{r}-\bm{r}^\prime|}d\bm{r}^\prime,
\end{eqnarray}
where $\bm{r}$ and $\bm{r}^\prime$ are the position vectors of the projectile and the bound state electrons with respect to the nucleus. $\psi_{l m}(\bm{r^\prime})$ is the wave function of the hole state.
As a result, the hole potential for the $m = 0$ and  $|m| = 1$ vacancy states in the Xe$^+$ is expressed as
\begin{equation}\label{hole} 
  \begin{cases}
            V_{1,0}(r) =\int \frac{|P_{5,1}(r^\prime)|^2 }{r_>}dr^\prime + \frac{2}{5}  \int |P_{5,1}(r^\prime)|^2  \frac{(r_<)^2}{(r_>)^3}dr^\prime \\
           V_{1,1}(r) =\int \frac{|P_{5,1}(r^\prime)|^2}{r_>}dr^\prime - \frac{1}{5}   \int |P_{5,1}(r^\prime)|^2 \frac{(r_<)^2}{(r_>)^3}dr^\prime.
         \end{cases}
\end{equation}
where $P_{n,\ell}(r^\prime)= rR_{n,\ell}(r^\prime)$ is the radial wave function, for $Xe^+$ with $n,\ell =5,1$. Furthermore, $r_< = \mathrm{min}(r,r^\prime)$ ($r_> =\mathrm{max}(r,r^\prime) $) which represents the smaller (larger) value of $r$ or $r^\prime$.

\subsection{The QRS model for HATI}

According to the QRS theory~\cite{QRS_Ar, QRS, Fang}, the detected photoelectron momentum distributions can be factorized as a product of the momentum distribution
of the returning wave packet (RWP) and the differential cross section (DCS) for elastic scattering of the returning electron from the parent ion. By defining the HATI photoelectron momentum
distribution obtained from the SFA as
\begin{eqnarray} \label{HATI}
    D^\mathrm{\,HATI}_\mathrm{\,SFA2} (\bm{p}) = |f_\mathrm{\,SFA2} (\bm{p})|^2,
\end{eqnarray}
the QRS model for HATI reads~\cite{QRS_HATI2},
\begin{eqnarray} \label{HATI_QRS}
D^\mathrm{\,HATI}_\mathrm{\,QRS} (p, \theta) = W_\mathrm{\,SFA2}(p_r) \frac{d \sigma
^\mathrm{\,el} (p_r,\theta_r)}{d \Omega_r},
\end{eqnarray}
where $d \sigma
^\mathrm{\,el} (p_r,\theta_r)/d \Omega_r$ is the DCS for elastic scattering of the returning electron with the parent ion obtained from Eq.~(\ref{DCS}). $W_\mathrm{\, SFA2}(p_r)$ is the RWP describing the momentum distribution of the returning electron, which can be obtained by 
\begin{eqnarray} \label{HATI_SFA}
W_\mathrm{\,SFA2}(p_r) = D^\mathrm{\,HATI}_\mathrm{\,SFA2} (p, \theta) /  \frac{d \sigma_\mathrm{PWBA}^\mathrm{\,el} (p_r,\theta_r)}{d \Omega_r},
\end{eqnarray}
and is independent of the rescattering angle \( \theta_r \). We make the common choice of a large scattering angle \( \theta_r = 178^\circ \)~\cite{QRS_Ar}. Here $d \sigma_\mathrm{PWBA}^\mathrm{\,el} (p_r,\theta_r)/d \Omega_r$ is evaluated using the plane-wave first-order Born approximation. And $p$, $p_r$, $\theta$ and $\theta_r$ are the detected momentum, rescattering momentum, detected angle, and rescattering angle, respectively.

The detected momentum $\bm{p}$ and rescattering momentum $\bm{p}_r$ are related by
\begin{eqnarray} \label{pr_Ar}
\bm{p}=\bm{p}_r-\bm{A}_r,
\end{eqnarray}
where the additional momentum $\bm{A}_r$ is the vector potential of the laser field at the recollision time. We use the approximation 
\begin{equation}
    A_r=p_r/1.26, \label{Ar_pr}
\end{equation}
and this relation is determined approximately by solving Newton’s equation of motion for an electron in a monochromatic laser field~\cite{QRS_Ar}. As a result, the momentum distribution $D^{\text{HATI}}(p,\theta)$ can be understood as a superposition of circles with radii $p_r$ and centers $A_r$. Tracing the angular distribution on these circles gives access to the DCSs. 

 
\section{Results and discussion}\label{Sec_Results}

\begin{figure}[b]
\includegraphics[width=0.54\textwidth,angle=270]{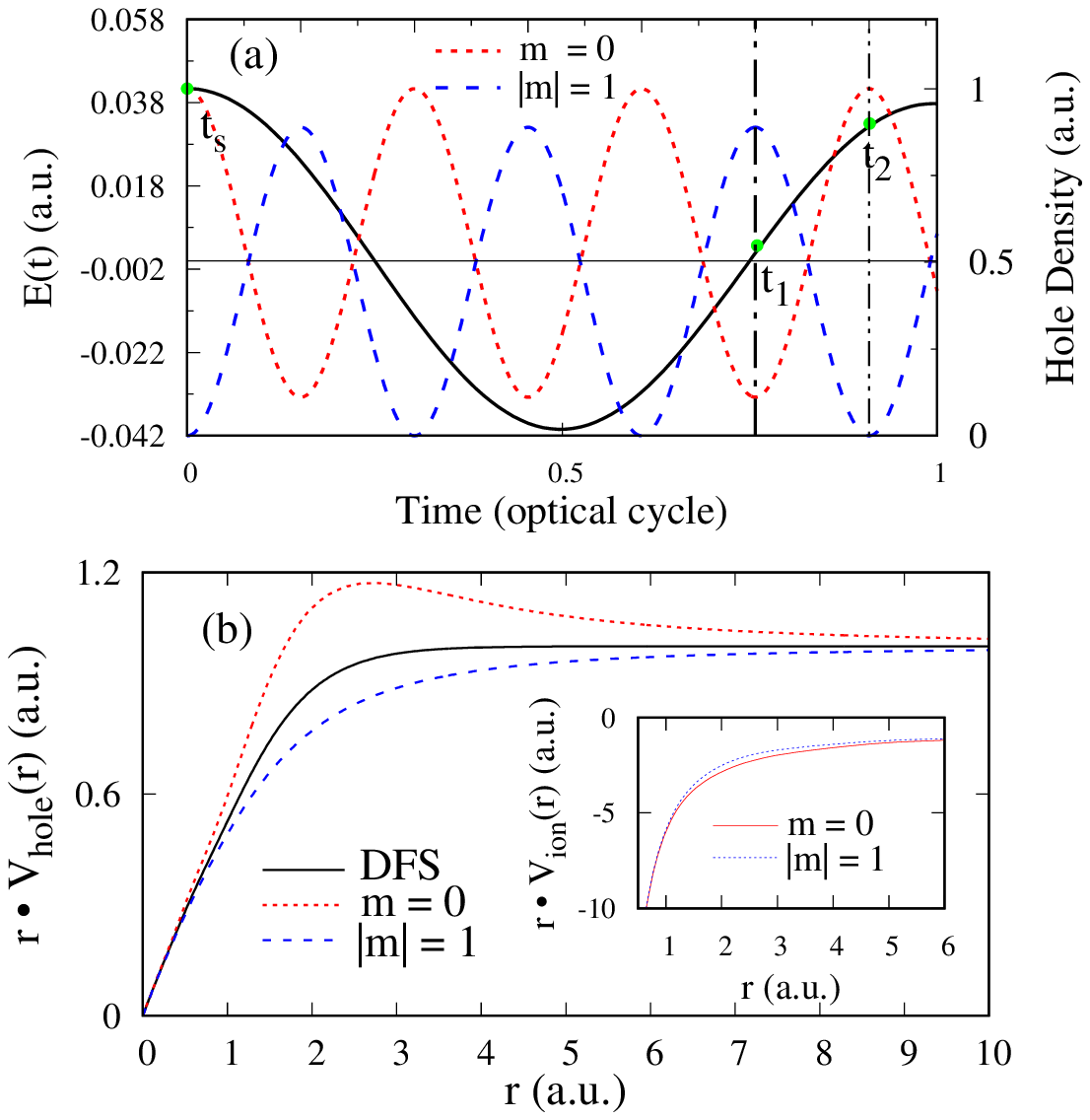}
\caption{\label{fig_1} (Color online)  (a) Schematic of the continuum and bound electron dynamics induced by tunnel ionization. The oscillation of the laser electric field (solid black curve) is compared to the hole population for the $m = 0$ (dotted red curve) and  $|m| = 1$ (dotted blue curve) vacancy states. The times $t_1 = 2.5\,T_{\mathrm{SO}}$ and $t_2 = 3.0 \, T_{\mathrm{SO}}$ mark times at which the hole populates dominantly the $|m|=1$ and $m=0$ states, respectively. (b) Electron-hole potentials ($V_{10}$ and $V_{11}$ for the $m = 0$ and $|m|= 1$ vacancy states of Xe$^+$, and the Dirac-Fock-Slater potential~\cite{JAC} representing the mean field of all electrons. The inset shows the full scattering potential of the ion, i.e.~includng the Coulomb term of equation \ref{holepotential}.}
\label{fig:1}
\end{figure}

Figure \ref{fig:1}(a) illustrates the time-evolution of the electric field $E(t)$ and the hole-state density in the Xe$^+$ ion in our experiment. Near the peak of the laser field, around $t_s$, an electron tunnels from the atom and is subsequently accelerated in the laser field. According to the classical recollision model, the electron returns to the parent ion at a time $t_{1,2}$, roughly $3/4 \pm 1/4$ of an optical cycle after emission, and carries a momentum $p_r(t_r) = -\left(A(t_r-A(t_s)\right)$, where $A(t)$ is the vector potential of the laser field. The travel time $\Delta t= (t_r - t_s)$ of the returning electron corresponds to the delay between electron emission (pump) and recollision (probe). The second clock, corresponding to the spin-orbit wavepacket motion, is also started at $t_s$. Given the period $T_{\mathrm{SO}}=\SI{3.2}{fs}$ in Xe$^+$, it is desirable to probe the wave packet at times spanning over $\SI{1.6}{fs}$ apart. With the wavelength of $\SI{3100}{nm}$ (optical period $T = \SI{10.5}{fs}$), we identify the recollision times $t_1$ and $t_2$, corresponding to delays of $\Delta t_1= 2.5 T_{\mathrm{SO}} = \SI{8.0}{fs}$ and $\Delta t_2 = 3.0 T_{\mathrm{SO}} = \SI{9.6}{fs}$ at which the hole is expected to populate primarily the the $|m| = 1$ ($m = 0$) states, respectively.

Figure~\ref{fig:1} (b) presents the electron-hole potentials weighted by the radial distance for Xe$^+$ as well for comparison. It can be seen from Fig.~\ref{fig:1} (b) that the potentials for $m = 0$ and $|m |= 1$ hole states, as well as the Dirac-Fock-Slater potential have the same asymptotic behavior at $r = \infty$. However, as in Eq.~(\ref{hole}), it is noticed that the potentials of the $m = 0$ and $|m| = 1$ orbitals differ significantly at around $r = 2$. A similar trend can also be observed for the constructed ion potential in the inset of Fig.~\ref{fig:1} (b), where the ion potential with the $|m| = 1$ orbital is slightly larger than the ion potential for the $m = 0$ hole state around $r = 2$.


\begin{figure}[htbp]
\includegraphics[width=0.37\textwidth,angle=270]{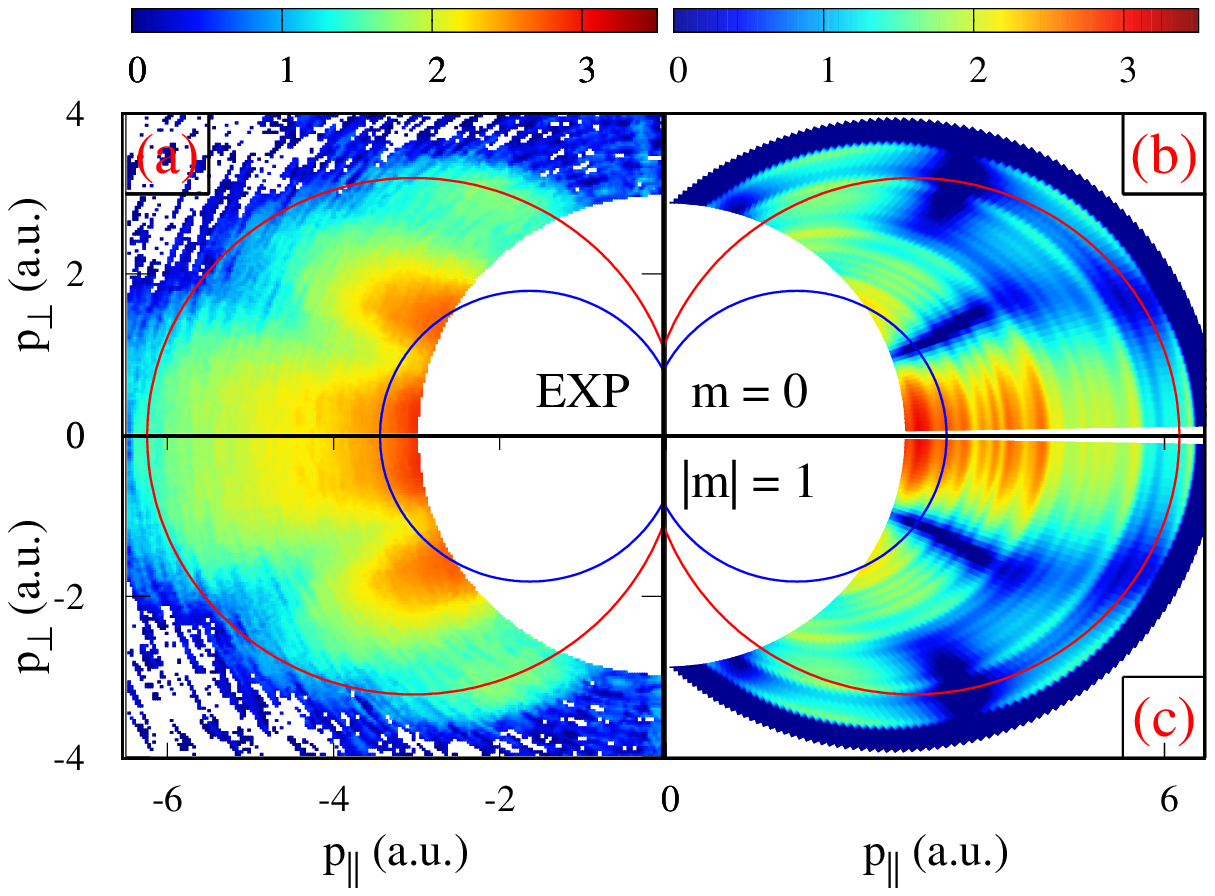}
\caption{Left: Measured photoelectron momentum distribution for HATI of xenon using 40-fs laser pulses with a central wavelength of $\SI{3100}{nm}$ and a peak laser intensity of $6\times10^{13}$ W/cm$^2$. Right: Results of the QRS calculations for the $m = 0$ (up row) and $|m|= 1$ (below row). The red and blue circles correspond to trajectories with travel times corresponding to $2.5\, T_\mathrm{SO}$ ($p_r = \SI{1.8}{\atomicunit}$) and $3.0\, T_\mathrm{SO}$ ($p_r = \SI{3.4}{\atomicunit}$). The central region of the momentum distribution, which is dominated by direct electrons, is removed in order to improve the visibility of the momentum distribution of rescattered electrons.}
\label{Fig2_PMD} 
\end{figure}

Experiments have been carried out using the mid-infrared (MIR) laser \cite{Kurucz2020} at the ELI-ALPS laser facility in Szeged, Hungary. The laser provides 40\,fs pulses centered around $\lambda = \SI{3100}{nm}$ at a repetition rate of \SI{100}{kHz}. A pair of wire grid polarizers are used to obtain linearly polarized light with adjustable power. The polarization direction is subsequently adjusted using a motorized broadband half-wave plate (B.~Halle). The laser pulses are sent into a stereographic photoelectron time-of-flight spectrometer \cite{Kuebel2021}. The laser is back-focused ($f = \SI{10}{cm}$) in front of an effusive nozzle injecting Xe gas into the vacuum chamber. Photoelectrons created in the laser focus are detected within a narrow solid angle ($\approx. 3^\circ$) using microchannel plate detectors mounted at a distance of \SI{50}{cm}, on either side of the spectrometer. Measurements of the photoelectron momentum distribution in the polarization plane are sampled by rotating the polarization axis of the laser and collecting time-of-flight spectra at each angle. The experimental results presented below are symmetrized with respect to reflection at the $p_\perp = 0$ axis. Small asymmetries observed in the raw data indicate a slight ellipticity introduced by the half-wave plate used in the experiment.

In Fig.~\ref{Fig2_PMD}, we compare the photoelectron momentum distributions for laser-induced ionization and scattering from Xe obtained experimentally as shown in Fig.~\ref{Fig2_PMD}(a) with the results of our modeling presented in Fig.~\ref{Fig2_PMD}(b) and (c). The experimental data exhibits pronounced modulations in the angular distribution of the photoelectrons. These are well reproduced by the QRS results. Despite some discrepancies regarding the electron yield, the good qualitative agreement between experimental and theoretical data along the polarization axis indicates that the cross-sections used here are suitable for describing laser-induced rescattering from Xe. However, we found out that the simulated momentum distributions for the $m = 0$ and $|m|=1$ vacancy states are very similar to each other. These small differences cannot be discerned in the experimental data. 

\begin{figure}[htbp]
\includegraphics[width=0.47\textwidth,angle=270]{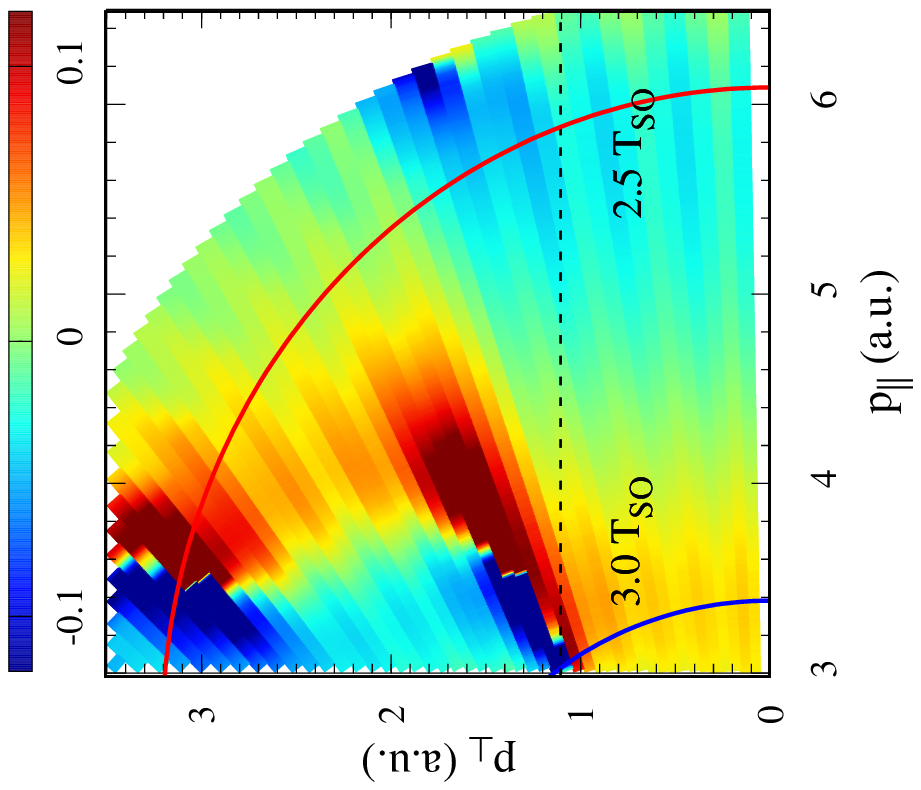}
\caption{\label{fig3_A} The normalized difference between the calculated PMD for the recollision of $m = 0$ and $|m| = 1$ states in the Xe$^+$ ions. The red and blue circular segments correspond to trajectories with travel times corresponding to $2.5\, T_\mathrm{SO}$ ($p_r = \SI{1.8}{\atomicunit}$) and $3.0\, T_\mathrm{SO}$ ($p_r = \SI{3.4}{\atomicunit}$).}
\end{figure} 

Since the computational results for the two hole states are nearly indistinguishable when viewed on the log scale, it is instructive to represent in Fig.~\ref{fig3_A} the normalized difference of these spectra, which is defined as
\begin{eqnarray}
\label{A}
A=\frac{D_{m=0}-D_{|m|=1}}{D_{m=0}+D_{|m|=1}},
\end{eqnarray}
where $D_{m=0}$ ($D_{|m|=1}$) are the momentum distributions calculated for $m=0$ ($|m| = 1$). In this way, we extract the differences in the momentum distributions arising from electron scattering from the $m = 0$ and $|m| = 1$ vacancy states.  The maxima and minima in this normalized difference plot provide information on where the photoelectron momentum distributions provide contrast between the two vacancy states. Specifically, for small values of $p_\perp$, the maximum positive contrast (more signal for $m = 0$) is obtained around the final momenta of $p_{||}=\SI{3.5}{\atomicunit}$, while the maximum negative contrast (more signal for $|m| = 1$) is observed at the final momenta of $p_{||}=\SI{6.0}{\atomicunit}$. These values coincide with the scattering rings which correspond to the maximum contrast in the population density, indicated as red and blue rings in Figs.~\ref{Fig2_PMD} and \ref{fig3_A}, respectively. The numerical results demonstrate that the electron signals due to rescattering from the $m=0$ and $|m|=1$ hole states are, in principle, distinguishable. However, the direct measurement of the normalized difference, as presented in Fig.~\ref{fig3_A}, is not straight forward. It would require usage of a combination of different laser wavelengths and accurately chosen intensities. Additionally, one could exploit the fact that the spin-orbit period for Kr (\SI{6.2}{fs}) is twice as long as for Xe. 

\begin{figure*}[htbp]
\includegraphics[width=0.29\textwidth,angle=270]{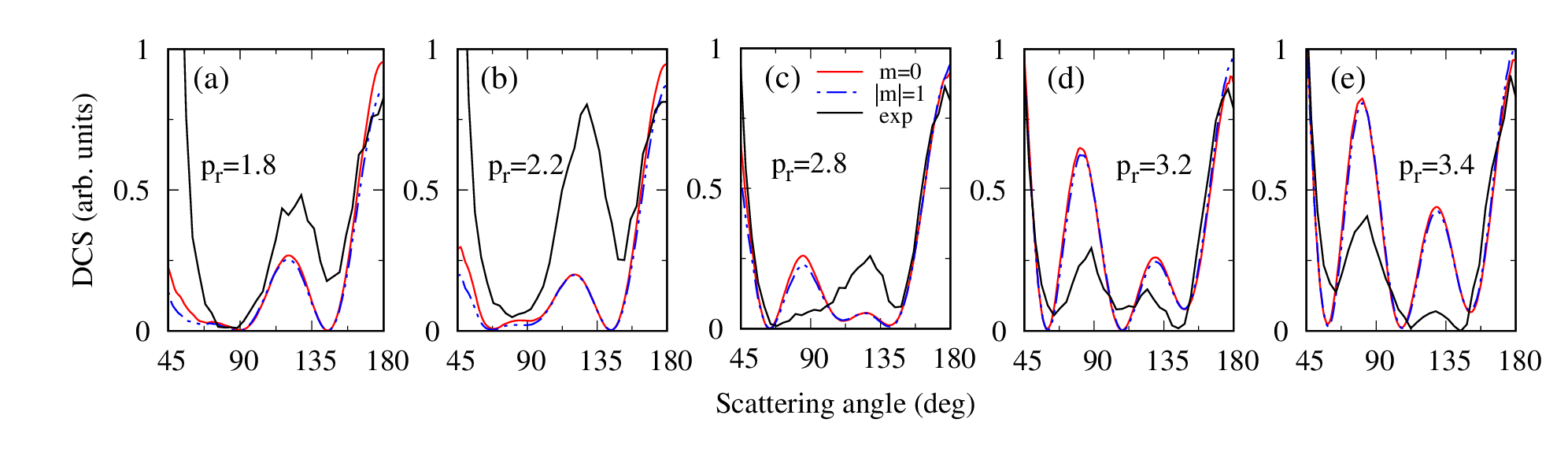}
\caption{\label{fig4_DCS} DCSs for elastic electron scattering from Xe$^{+}$ in the $m = 0$ (solid red curve) and $|m| = 1$ (dashed blue curve) hole states for recollision momenta of (a) $\SI{1.8}{\atomicunit}$, (b) $\SI{2.2}{\atomicunit}$, (c) $\SI{2.6}{\atomicunit}$, (d) $\SI{2.8}{\atomicunit}$, (e) $\SI{3.2}{\atomicunit}$, respectively. The theoretical DCS values are compared to experimental values extracted from the measured photoelectron angular distributions.}
\end{figure*}

 For a quantitative analysis, we present the calculated DCS values for the two hole states in Fig.~\ref{fig4_DCS}, and compare them to the values extracted from the experimental results at various recollision momenta. 
 The differences observed in Fig.~\ref{fig3_A} are reflected in the DCS. For example, at low momenta $p_r\sim\SI{2}{\atomicunit}$, the DCS at $180^\circ$ is larger for $m = 0$ (cf.~red signal in Fig.~\ref{fig3_A}). 
 At higher momenta, $p_r \sim \SI{3}{\atomicunit}$, however, the DCS at $180^\circ$ is larger for $|m| = 1$. 
 While the comparison of the momentum distributions in Fig.~\ref{Fig2_PMD} indicates a good qualitative agreement between the angular distributions observed in the experimental and numerical results, it is evident here that the agreement does not reach the quantitative level necessary to distinguish between the subtle differences observed in the theoretical results. 
 The reasons for the discrepancies between the measured and calculated angular distribution may include several experimental factors, such as the slightly elliptical polarization mentioned above. Moreover, the experimental results are subject to averaging over the spatial intensity distribution in the focus, and the temporal variations of the instantaneous intensity throughout the laser pulse. These effects are not taken into account in the numerical simulations, which are based on the SFA and QRS models. The QRS model used here approximates the temporal variations of the rescattering momentum and vector potential by Eq.~(\ref{Ar_pr}), which affects the resulting momentum distribution. 

\section{Conclusions and Outlook}\label{Sec_Conclusions}

We present a study on the interplay between bound and continuum electron dynamics initiated by strong-field ionization of xenon. Specifically, we investigate whether the ensuing spin-orbit electron dynamics can be probed through laser-induced electron diffraction. The two-dimensional photoelectron momentum distributions for HATI of Xe are calculated for rescattering from the $m=0$ and $|m|=1$ hole states, using the QRS theory. This work represents an initial attempt to experimentally explore valence electron dynamics Xe through LIED. While the numerical results agree with the experimental ones on a qualitative level, they do not reach the quantitative level necessary to distinguish between the rescattering signal from the two hole states. Addressing this challenge likely requires advanced theoretical approaches and more accurate experimental data. If such data becomes available in the future, an artificial intelligence approach may aid the interpretation of the data and enable the observation of valence electron dynamics by LIED. This intriguing problem underscores the need for future research to address the complex issue of spin-orbit effects during LIED.

\section*{Acknowledgements}
The authors acknowledge fruitful discussions with S. Carlstr\"om, J.M. Dahlstr\"om and S. Patchkovskii. 
This work was supported by the Deutsche Forschungsgemeinschaft (DFG, German Research Foundation) under Project No.\ 440556973 and the Emmy Noether program, project No. 437321733.

\end{document}